\documentclass[fleqn,twoside]{article}
\usepackage{espcrc2}

\usepackage{graphicx}
\usepackage[figuresright]{rotating}
\usepackage[dvips,usenames]{color}

\newcommand{\AmS}{{\protect\the\textfont2
A\kern-.1667em\lower.5ex\hbox{M}\kern-.125emS}}

\title{Bethe-Salpeter equation in Minkowski space with cross-ladder kernel}

\author{V.A. Karmanov\address
{Lebedev Physical Institute,
\\
Leninsky Prospekt 53, 119991 Moscow, Russia}%
        \thanks{Supported in part by the RFBR grants  05-02-17482-a
        and 05-02-26615}
        and
        J. Carbonell\address{Laboratoire de Physique Subatomique
        et Cosmologie,
        \\
53 avenue des Martyrs, 38026 Grenoble, France}
}

\begin{document}

\begin{abstract}
A new method for solving the Bethe-Salpeter equation is developed.
It allows to find the Bethe-Salpeter amplitudes both in Minkowski
and in Euclidean spaces and, as a by product, the light-front wave
function. The method is valid for any kernel given by irreducible
Feynman graphs. Bethe-Salpeter and Light-Front equations for
scalar particles with ladder + cross-ladder kernel are solved.
 \vspace{1pc}
\end{abstract}

\maketitle

\section{INTRODUCTION}\label{intr}

Bethe-Salpeter (BS) equation \cite{SB_PR84_51}  is an important tool for studying the
relativistic  bound state problem in a field theory framework (see
for review~\cite{nakanishi}). For a bound state with total
momentum $p$ and in case of equal mass particles, it reads
\begin{equation}\label{bs}
\Phi(k,p)= G_0^{(12)}\int
\frac{d^4k'}{(2\pi)^4}iK(k,k',p)\Phi(k',p)\ ,
\end{equation}
where $\Phi$ is the BS amplitude, $K$ the interaction kernel, $m$
the mass of the constituents, $k$ their relative momentum and
$G_0^{(12)}$  the free propagators of the constituent particles:
\begin{eqnarray}\label{G0}
&&G_0^{(12)}(p,k) =G_0^{(1)}(p,k)G_0^{(2)}(p,k)=
\\
&&=\frac{i}{(\frac{p}{2}+ k)^2-m^2+i\epsilon} \;
\frac{i}{(\frac{p}{2}-k)^2-m^2+i\epsilon}\ . \nonumber
\end{eqnarray}
We will denote by $M=\sqrt{p^2}$ the total mass of the bound
state, and by $B=2m-M$ its binding energy.

The  BS equation has singularities which make difficult to find
its numerical solution. These singularities are due to the free
propagators (\ref{G0}) but can also result from the interaction
kernel itself. To overcome this difficulty, Wick \cite{WICK_54}
formulated the BS equation in the Euclidean space, by  rotating
the relative energy $k_0$ in the complex plane, {\it i.e.}, by
introducing the variable $k_4=-ik_0$. This leads to a well defined
integral equation -- see  (\ref{bs2}) below -- which can be solved
by standard methods and provides the mass of the system. In this
procedure the original BS amplitude is however lost, and the
"rotated" one can no longer be used in calculating other physical
observables, like for instance form factors.

A successful attempt to calculate the BS amplitude in Minkowski
space for the ladder kernel was presented in \cite{KW}. It was
based on the Nakanishi integral representation of the amplitude
\cite{nak63}. Another approach to solve BS equation with separable
interaction in Minkowski space was developed in~\cite{bbmst} and
applied to the nucleon-nucleon system. In \cite{sfcs}, an equation
obtained by projecting the BS equation on the light-front (LF)
plane was studied: the LF kernel was found approximately, as an
expansion in terms of the BS one and the original BS amplitude was
not reconstructed from its LF projection.

We present in this paper -- based on our works \cite{bs1,bs2} --
a new method for solving the BS equation. This method provides the
BS amplitude both in Minkowski and in Euclidean space, depending
on the value (real or imaginary) of the $k_0$ variable. Our
approach is based on a projection of the BS equation on the LF
plane, and on the Nakanishi integral representation \cite{nak63}
of the BS amplitude. The LF projection plays the role of an
integral transform which removes the singularities of the BS
amplitude. The Nakanishi representation results in an equation in
a closed form and allows to easily restore with its solution, the
original BS amplitude. The transformed equation is derived without
any approximation and remains equivalent to the original BS one.
Our method is not restricted to the ladder kernel. For more
complicated interactions, {\it e.g.} cross box, calculations
become more lengthy, but the additional difficulties are due to
the evaluation of the Feynman diagram for the kernel itself  and
not to the solution of the equation.

In order to present the method more distinctly, we consider the
case of zero total angular momentum and spinless particles.

In sect. \ref{project} we derive the equation allowing to find the
BS amplitude both in Minkowski and Euclidean spaces. In sect.
\ref{LFD}  we present corresponding LF equation and in sect.
\ref{espace} -- the BS equation in Euclidean space. In sect.
\ref{num} the numerical solutions  are presented. Sect.
\ref{concl} contains some concluding remarks.

\section{BETHE-SALPETER EQUATION ON THE LIGHT-FRONT PLANE}\label{project}

Our method is inspired by an existing relation  -- see eq.
(\ref{lfwf3a}) below -- between the (singular) BS amplitude
$\Phi(k,p)$ and the (non-singular) two-body LF wave function
$\psi(\vec{k}_{\perp},x)$. The latter can be obtained by
projecting the BS amplitude on the LF plane. We use the covariant
formulation \cite{cdkm} with the LF plane, defined by $\omega\cdot
x=0$ with $\omega^2=0$. The particular choice
$\omega=(\omega_0,\vec{\omega})=(1,0,0,-1)$ results in the
standard LF form $t+z=0$. We apply to both sides of (\ref{bs}) the
integral transform:
\begin{eqnarray}\label{bsproject}
&&\!\!\!\!\!\!\!\!\!\int_{-\infty}^{\infty}d\beta\,\Phi(k+\beta\omega,p)
= \int_{-\infty}^{\infty}d\beta \,G_0^{(12)}(k+\beta\omega,p)
\nonumber\\
&&\!\!\!\!\!\!\!\!\!\times \int
\frac{d^4k'}{(2\pi)^4}iK(k+\beta\omega,k',p)\Phi(k',p)\ ,
\end{eqnarray}

In the transformed equation, the BS amplitude is written in terms
of the Nakanishi integral representation~\cite{nakanishi}, which
for zero angular momentum reads:
\begin{eqnarray}\label{bsint}
&&\Phi(k,p)=\frac{-i}{\sqrt{4\pi}}\int_{-1}^1dz'\int_0^{\infty}d\gamma'
\\
&&\times \frac{g(\gamma',z')}{\left[\gamma'+m^2
-\frac{1}{4}M^2-k^2-p\cdot k\; z'-i\epsilon\right]^3}\ .
 \nonumber
\end{eqnarray}
A similar representation exists for non-zero angular momentum.
This representation is valid for a rather wide class of solutions.
This leads to the following equation, derived in \cite{bs1}, for
the weight function $g(\gamma,z)$:
\begin{eqnarray}
\label{bsnew} &&\int_0^{\infty}\frac{g(\gamma',z)d\gamma'}
{\Bigl[\gamma'+\gamma +z^2 m^2+(1-z^2)\kappa^2\Bigr]^2}=
\nonumber\\
&& \int_0^{\infty}d\gamma'\int_{-1}^{1}dz'\;V(\gamma,z;\gamma',z')
g(\gamma',z')\ ,
\end{eqnarray}
where $V$ is a kernel given in terms of the BS interaction kernel
$K$ by
\begin{eqnarray}\label{V}
&&V(\gamma,z;\gamma',z')=  \\
&&
 \int_{-\infty}^{\infty}d\beta\;
G_0^{(1)}(p,k+\beta\omega)G_0^{(2)}(p,k+\beta\omega)
\nonumber\\
&&\times\frac{\omega\cdot p}{2^4\pi^5}\int\frac{K(k+\beta
\omega,k',p)\,d^4k'} {\left[{k'}^2+p\cdot k' z'-\gamma'-\kappa^2
+i\epsilon\right]^3}\ .
 \nonumber
\end{eqnarray}
$G_0^{(i)}$ are the free propagators (\ref{G0}),
the denominator arises from the representation
(\ref{bsint}), the integration over $k'$ results from the BS equation
(\ref{bs}) and the integration over $\beta$ from the LF projection
(\ref{bsproject}). The bound state mass $M$ enters through the
parameter $ \kappa^2 = m^2- \frac{1}{4}M^2. $

The four-vectors in the r.h.s of eq. (\ref{V}) appear in form of
scalar products, expressed through the equation variables
$(\gamma,z)$ defined by:
\begin{eqnarray*}
\label{kin1} &&k^2=-\frac{(\gamma+z^2 m^2)}{1-z^2},\quad
\frac{\omega\cdot k}{\omega\cdot p}=-\frac{1}{2}z\ ,
\\
&&p\cdot k=\frac{z[\gamma+z^2 m^2+(1-z^2)\kappa^2]}{1-z^2},
\end{eqnarray*}
and related to the standard LF
variables by $\gamma=k_{\perp}^2$, $z=1-2x$\ .

Equation (\ref{bsnew}) is equivalent to the initial BS equation
(\ref{bs}) and provides,  for a given kernel $K$, the same
bound state mass $M$. Once $g(\gamma,z)$ is known, the BS
amplitude can be restored by eq. (\ref{bsint}). For real values of
$k_0$, we find the BS amplitude in Minkowski space and for
imaginary values $k_0=ik_4$, in the Euclidean one.
The corresponding LF wave function $\psi(k_\perp,x)$ is
easily obtained by
\begin{eqnarray} \label{lfwf3a}
&&\psi(k_\perp,x)=
\\
&& \frac{x(1-x)}{\pi}\int_{-\infty}^{\infty}\Phi(k+\beta\omega,p)
(\omega\cdot p)\,d\beta=
 \nonumber\\
&&\frac{1}{\sqrt{4\pi}}\int_0^{\infty}\frac{x(1-x)g(\gamma',1-2x)d\gamma'}
{\Bigl[\gamma'+k_\perp^2 +m^2-x(1-x)M^2\Bigr]^2}\ .
 \nonumber
\end{eqnarray}


\subsection{Ladder kernel}\label{le}

Application to the ladder BS kernel
\begin{equation}
\label{ladder}
iK^{(L)}(k,k',p)=\frac{i(-ig)^2}{(k-k')^2-\mu^2+i\epsilon}
\end{equation}
allows to test our method. In this case, the kernel $V=V^{(L)}$ of
equation (\ref{bsnew}) is calculated analytically.

It becomes especially simple for $\mu=0$. This particular case
constitutes the Wick-Cutkosky model
\cite{WICK_54,CUTKOSKY_PR96_54}. We search the solution of
(\ref{bsnew}) in the form:
$$
g(\gamma,z)=\delta(\gamma)\; g(z)
$$
justified by possibility to find a function $g(z)$ which does not
depend on $\gamma$. The integration over $\gamma'$ in both sides
of equation (\ref{bsnew}) drops out, the $\gamma$-dependence of
the kernel $V$ is reduced to a factor which is the same in both
sides of equation and cancels. After that, the equation
(\ref{bsnew}) takes a simplified form
\begin{equation}
\label{bsnew0}
g(z)=\frac{\alpha}{2\pi}\int_{-1}^{1}dz'\;\tilde{V}(z,z')g(z')
\end{equation}
with $\alpha=g^2/(16\pi m^2)$ and
\begin{eqnarray*}\label{Kt}
\tilde{V}(z,z')&=&\frac{m^2}{m^2-\frac{1}{4}(1-{z'}^2)M^2}
\nonumber\\
&\times&\left\{
\begin{array}{ll}
\frac{(1-z)}{(1-z')},&\quad\mbox{if $-1\le z'\le z\le 1$}\\
\frac{(1+z)}{(1+z')},&\quad\mbox{if $-1\le z\le z' \le 1$}
\end{array}
\right.
\end{eqnarray*}
which exactly coincides with the Wick-Cutkosky equation
\cite{nakanishi,WICK_54,CUTKOSKY_PR96_54}.

Numerical results for non-zero $\mu$ are given in sect. \ref{num}.

\subsection{Cross ladder kernel}\label{clad}
\begin{figure}[ht!]
\begin{center}
\includegraphics[width=6cm]{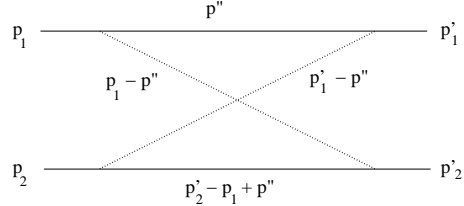}
\end{center}
\vspace{-1cm}
 \caption{Feynman cross graph.\label{CF}}
\end{figure}

The cross-ladder BS kernel is shown in Figure \ref{CF}. The
corresponding amplitude reads:
\begin{eqnarray}\label{F1}
&&K^{(CL)}(k,k',p)=\frac{-ig^4}{(2\pi)^4}
\\
&&\!\!\!\!\!\!\!\!\!\!\!\!\times\int
\frac{1}{[{p''}^2-m^2+i\epsilon] [(p'_2-p_1+p'')^2-m^2+i\epsilon]}
\nonumber\\
&&\!\!\!\!\!\!\!\!\!\!\!\!\times\frac{d^4p''}
{[(p_1-p'')^2-\mu^2+i\epsilon][(p'_1-p'')^2-\mu^2+i\epsilon]}\ .
\nonumber
\end{eqnarray}
We calculate this expression, substitute the result in (\ref{V}),
calculate the integrals and find in this way the cross-ladder
contribution to the kernel $V(\gamma,z;\gamma',z')$ in equation
(\ref{bsnew}). The full kernel is the sum of ladder and
cross-ladder graphs:
 $$
V=V^{(L)} +V^{(CL)}\ .
 $$
The details of calculation and expression for $V^{(CL)}$ are given
in \cite{bs2}.
%

\section{LIGHT-FRONT EQUATION AND KERNELS}\label{LFD}

We would like to compare the results obtained in the BS approach
with the ones found in Light-Front Dynamics (LFD). With this aim,
we precise here the LF equation and the corresponding kernel. In
terms of $\vec{k}_{\perp}$ and $x$  variables, the LF equation
reads (see {\it e.g.} \cite{cdkm}):
\begin{eqnarray}\label{eq1}
&&\!\!\!\!\!\left(\frac{\vec{k}^2_{\perp}+m^2}{x(1-x)}-M^2\right)
\psi(\vec{k}_{\perp},x)=-\frac{m^2}{2\pi^3}
\\
&&\!\!\!\!\!\times\int
V_{LF}(\vec{k}'_{\perp},x';\vec{k}_{\perp},x,M^2)
\frac{\psi(\vec{k}'_{\perp},x')d^2k'_{\perp}dx'}{2x'(1-x')}\ .
\nonumber
\end{eqnarray}

There  are two time-ordered ladder graphs and six LF time-ordered
cross-ladder graphs. The latters have the order $\alpha^2$. In
addition, and to the same order $\alpha^2$, there are two irreducible
time-ordered graphs with two mesons in the intermediate state
(stretched boxes). The full LFD kernel -- including ladder,
cross-ladder and stretched-box graphs -- is written in the form:
 $$
V_{LF}(\vec{k}'_{\perp},x';\vec{k}_{\perp},x,M^2)=V_{LF}^{(L)}+
V_{LF}^{(CL)}+V_{LF}^{(SB)}\ .
 $$
The LF equation (\ref{eq1}) has been solved with this full kernel.

\section{CROSS LADDER IN EUCLIDEAN SPACE}\label{espace}

The possibility of Wick rotation has been proved \cite{WICK_54}
for the ladder kernel. One can also "rotate", without crossing
singularities,  simultaneously all the energies $k_{i0}$ of a
perturbative Feynman amplitude. However, this possibility is not
evident for a BS amplitude with a kernel more complicated than the
ladder one, since now not all the energies are rotated: when the
relative energy $k_0$ is replaced by $ik_4$, the eigenvalue -- the
bound state energy $p_0$ -- still remains real. For the latter
case, the validity/invalidity of Wick rotation is discussed in the
literature. For the cross-ladder kernel, we can check it
numerically, by  finding the binding energies from  eq.
(\ref{bsnew}) -- see results on Table \ref{tab2} below -- and  by
solving the Euclidean space equation for L+CL kernel. One can then
see whether or not the Euclidean and Minkowski results coincide
with each other. Coincidence would confirm both: the validity of
Wick rotation for the cross ladder kernel and our method in
Minkowski space.

Equation in Euclidean space is obtained from (\ref{bs}) by the
replacement: $k_0=ik_4$, $k'_0=ik'_4$. In the rest frame
$\vec{p}=0$ it reads:
\begin{eqnarray}\label{bs2}
&&
\!\!\!\!\!\left[\left(k_4^2+\vec{k}\,^2+m^2
-\frac{M^2}{4}\right)^2+M^2k_4^2\right]\Phi_E(k_4,k)
 \nonumber\\
&& \!\!\!\!\!\!\!= \int \frac{ dk'_4
d^3k'}{(2\pi)^4}K_E(k_4,\vec{k};k'_4,\vec{k'})\Phi_E(k'_4,k')\ ,
\end{eqnarray}
where
$K_E(k_4,\vec{k};k'_4,\vec{k'})=K(ik_4,\vec{k};ik'_4,\vec{k'})$
and $\Phi_E(k_4,k)=\Phi(ik_4,k)$. We take sum of ladder, eq.
(\ref{ladder}), and cross ladder, eq. (\ref{F1}), "convert" them
in Euclidean space and find in this way the kernel
$K_E=K_E^{(L)}+K_E^{(CL)}$ of the equation (\ref{bs2}).

\section{NUMERICAL RESULTS}\label{num}

It turns out that the discretized integral operator in l.h.-side
of eq. (\ref{bsnew}) has very small eigenvalues. They are
unphysical but make  the solution unstable. We have regularize it
by adding a small constant $\varepsilon$ to its diagonal part. This
procedure allows us to obtain stable eigenvalues in the interval
$\epsilon=10^{-4}\div 10^{-12}$.
%
\begin{table}[ht!]
\begin{center}
\caption{Coupling constant $\alpha=g^2/(16\pi m^2)$ as a function
of the binding energy $B$ for ladder kernel with $\mu=0.15$ and
$\mu=0.5$ obtained with $\epsilon=10^{-6}$.}
\label{tab1}       
\begin{tabular}{ccc}
\hline\noalign{\smallskip}
$B$ & $\alpha(\mu=0.15)$ &  $\alpha(\mu=0.50)$  \\
\noalign{\smallskip}\hline\noalign{\smallskip}
0.01   &   0.5716           & 1.440 \\
0.10   &   1.437            & 2.498 \\
0.20   &   2.100            & 3.251 \\
0.50   &   3.611            & 4.901 \\
1.00   &   5.315            & 6.712 \\
\noalign{\smallskip}\hline
\end{tabular}
\end{center}
\end{table}
\begin{figure*}[ht!]
\centering
\includegraphics[width=7cm]{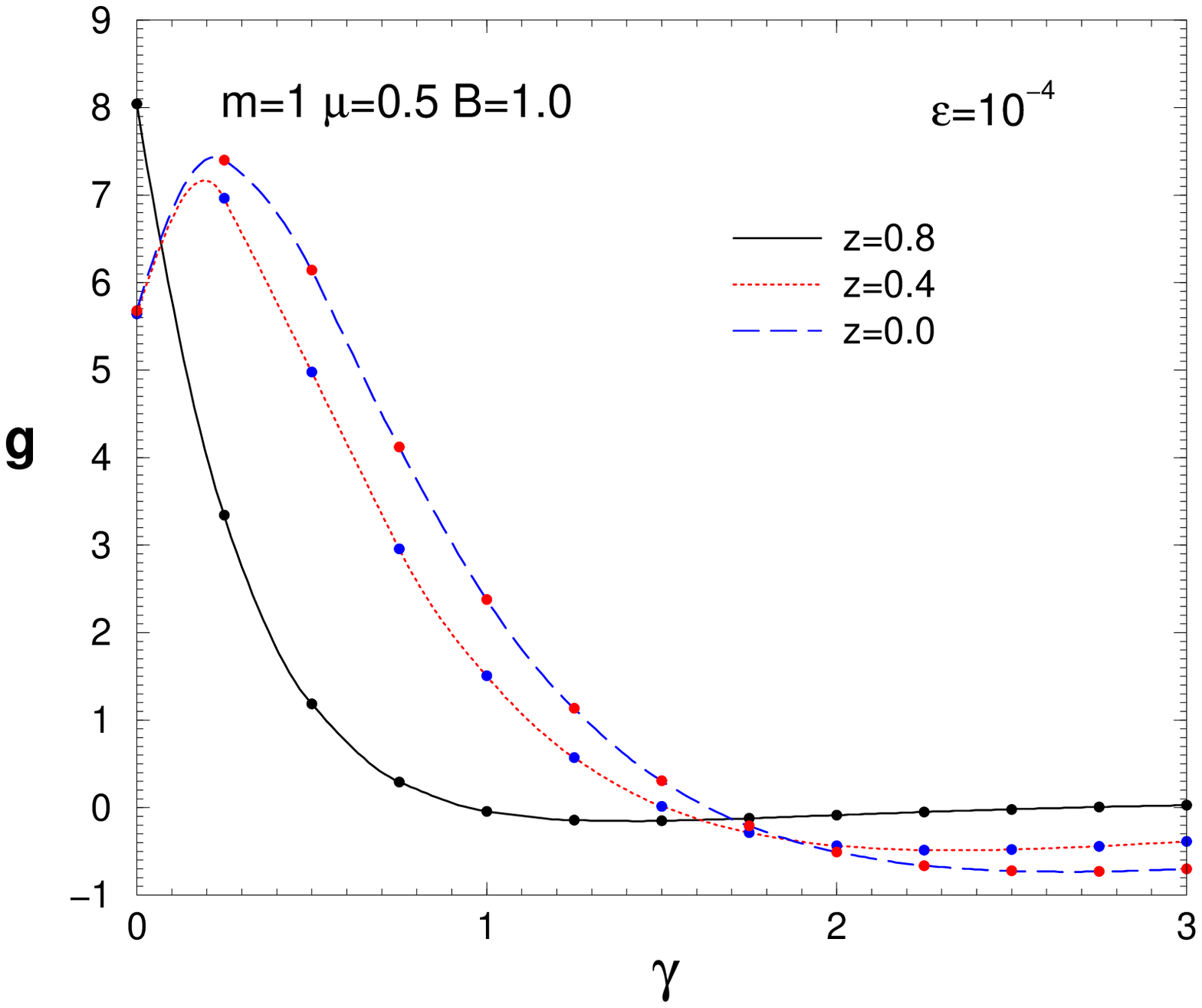}
\hspace{0.5cm}
\includegraphics[width=7cm]{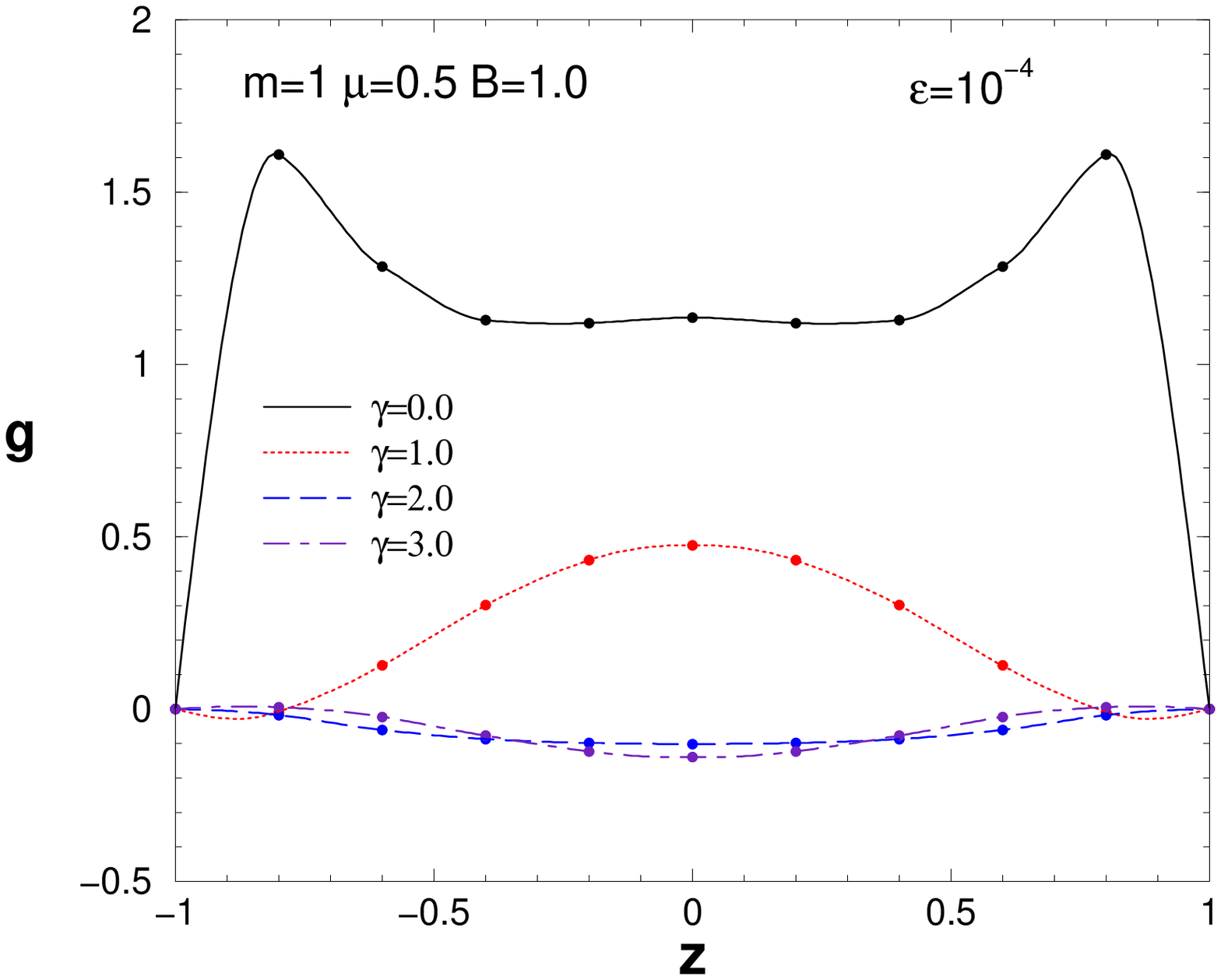}
\vspace{-1cm}
 \caption{Function $g(\gamma,z)$ for ladder kernel
for $\mu=0.5$ and $B=1.0$. On left -- versus $\gamma$ for fixed
values of $z$ and on right -- versus $z$ for a few fixed values of
$\gamma$.} \label{g_k}
\end{figure*}

For the ladder kernel with $\mu=0.15$ and $\mu=0.5$ and unit
constituent mass $m=1$ the values of binding energy $B=2m-M$ are
displayed  in Table \ref{tab1}. With all shown digits, they are in
full agreement with the results of \cite{mariane} obtained by
using the Wick rotation and the method of \cite{NT_FBS_96}. This
demonstrates the validity of our approach.

The corresponding  weight function $g$ for a system with $\mu=0.5$
and  $B=1.0$ is plotted in Figure \ref{g_k}.  Its
$\gamma$-dependence is not monotonous and has a nodal structure;
the $z$-variation  is also non trivial. We have remarked a strong
dependence
 of $g(\gamma,z)$ relative to values of the $\epsilon$
parameter smaller than $\sim10^{-4}$, in contrast to high
stability of corresponding eigenvalues. However, the corresponding
BS amplitude $\Phi$ and LF wave function $\psi$, obtained from
$g(\gamma,z)$ by the integrals (\ref{bsint}) and (\ref{lfwf3a}),
show the same strong stability as the eigenvalues.

\begin{figure*}[ht!]
\centering
\includegraphics[width=7cm]{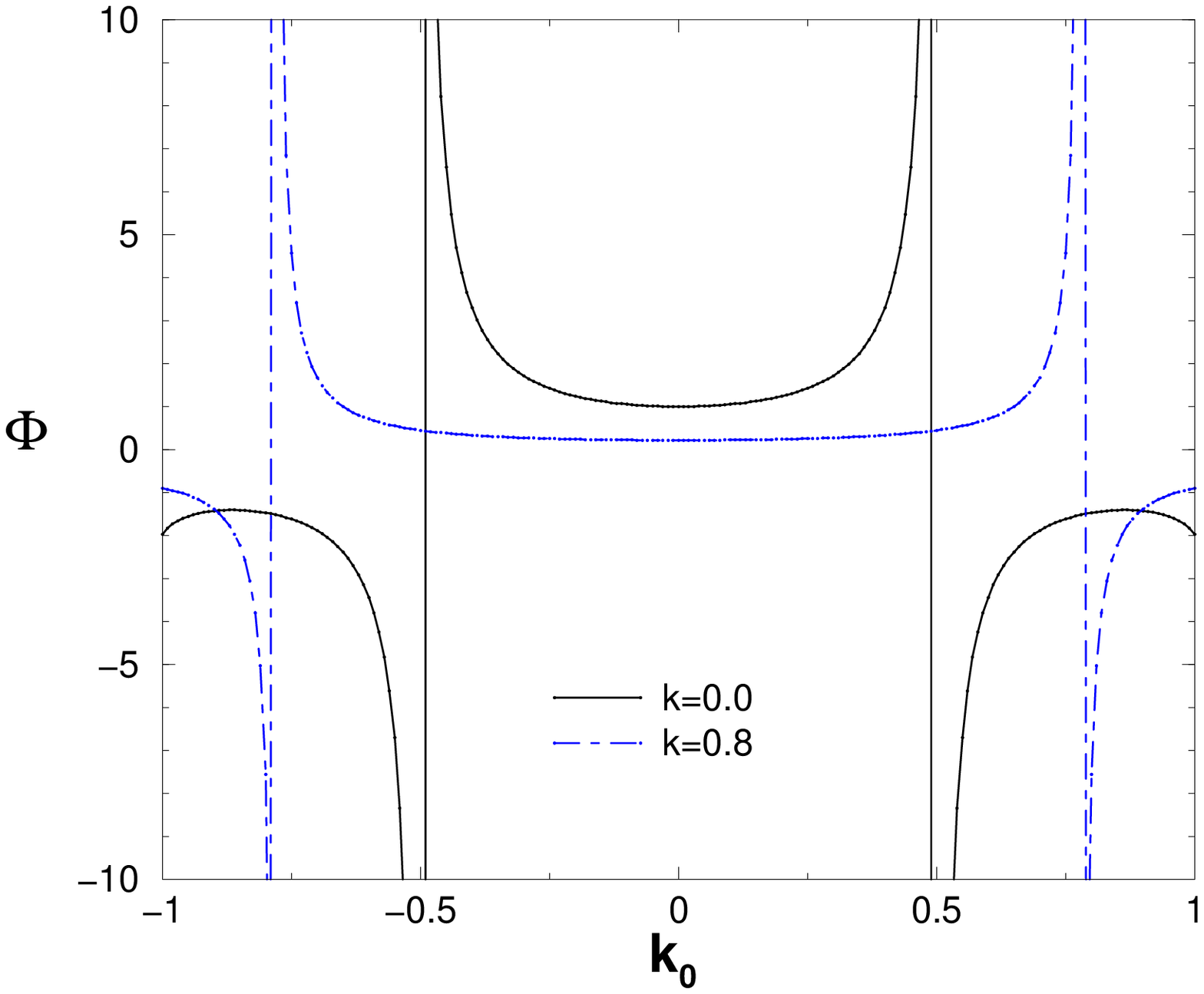}
\hspace{0.5cm}
\includegraphics[width=8cm]{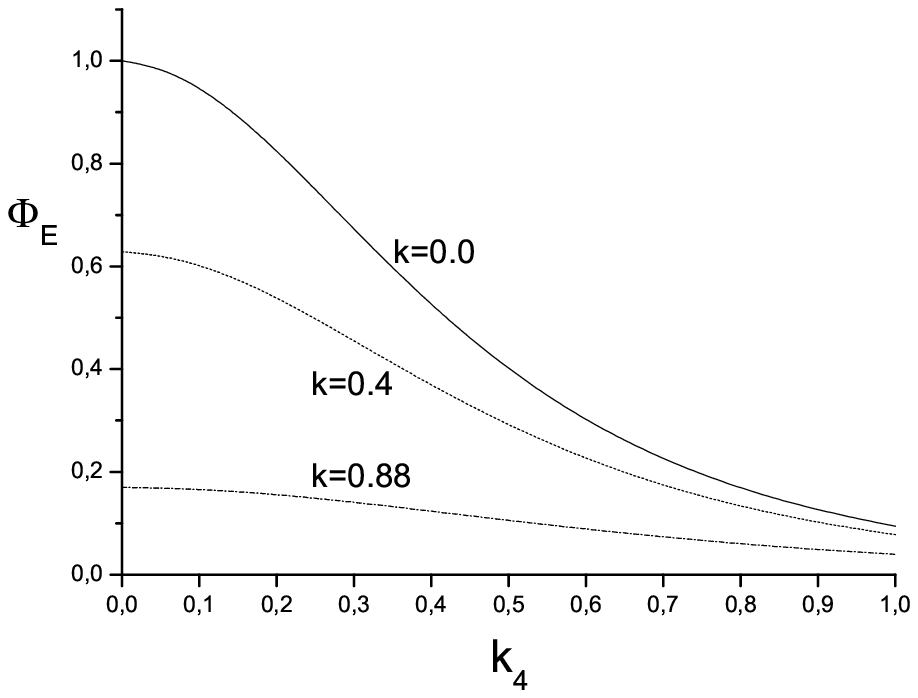}
 \vspace{-1cm}
 \caption{BS amplitude for ladder kernel with
$\mu=0.5$, $B=1.0$ and fixed values of $k$. On left,
$\Phi(k_0,k)$ in Minkowski space versus $k_0$; on right
$\Phi_E(k_4,k)=\Phi(k_0=ik_4,k)$ in Euclidean space versus $k_4$.}
\label{Phi_k}
\end{figure*}

\begin{figure*}[ht!]
\centering
\includegraphics[width=7cm]{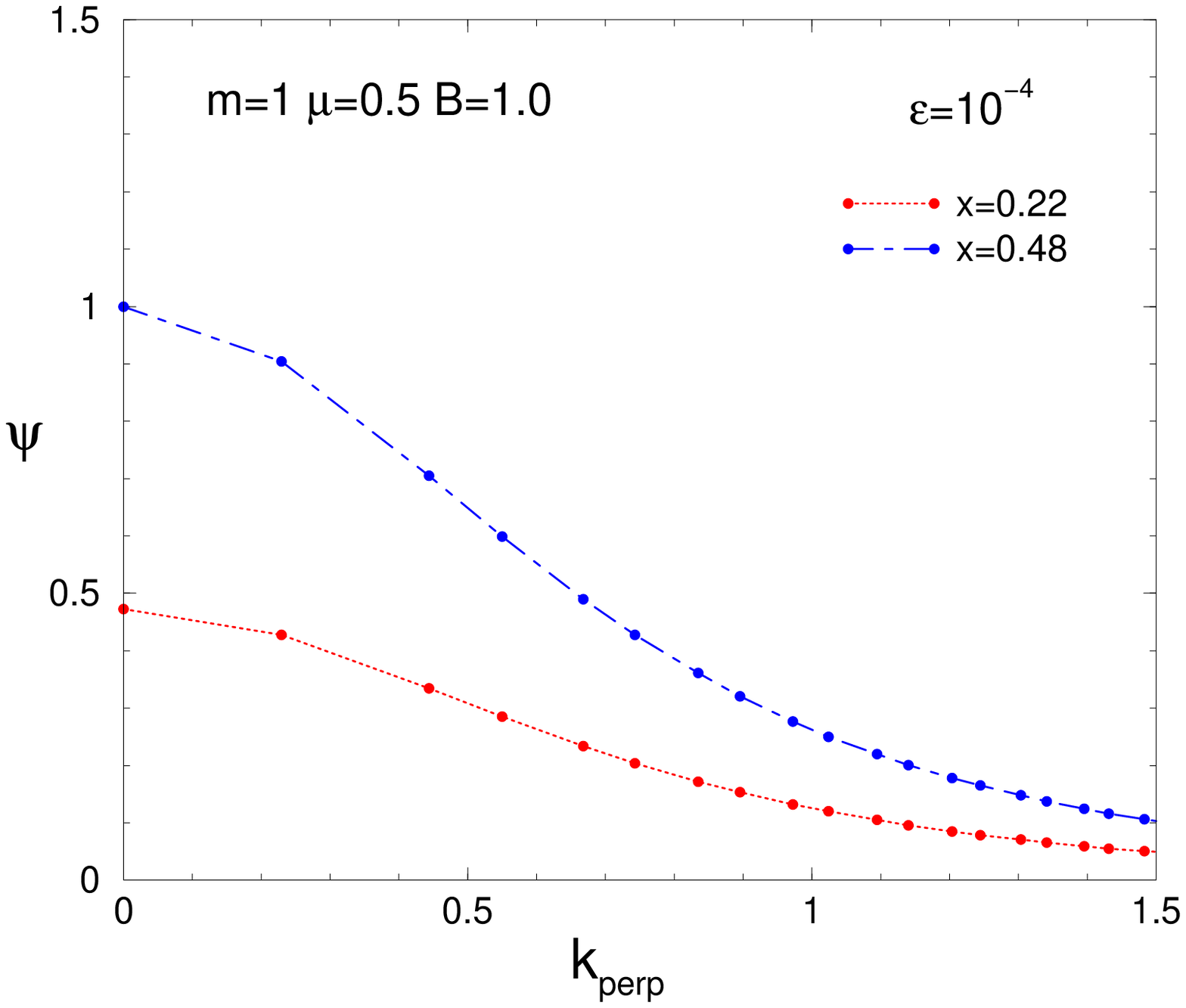}
\hspace{0.5cm}
\includegraphics[width=7cm]{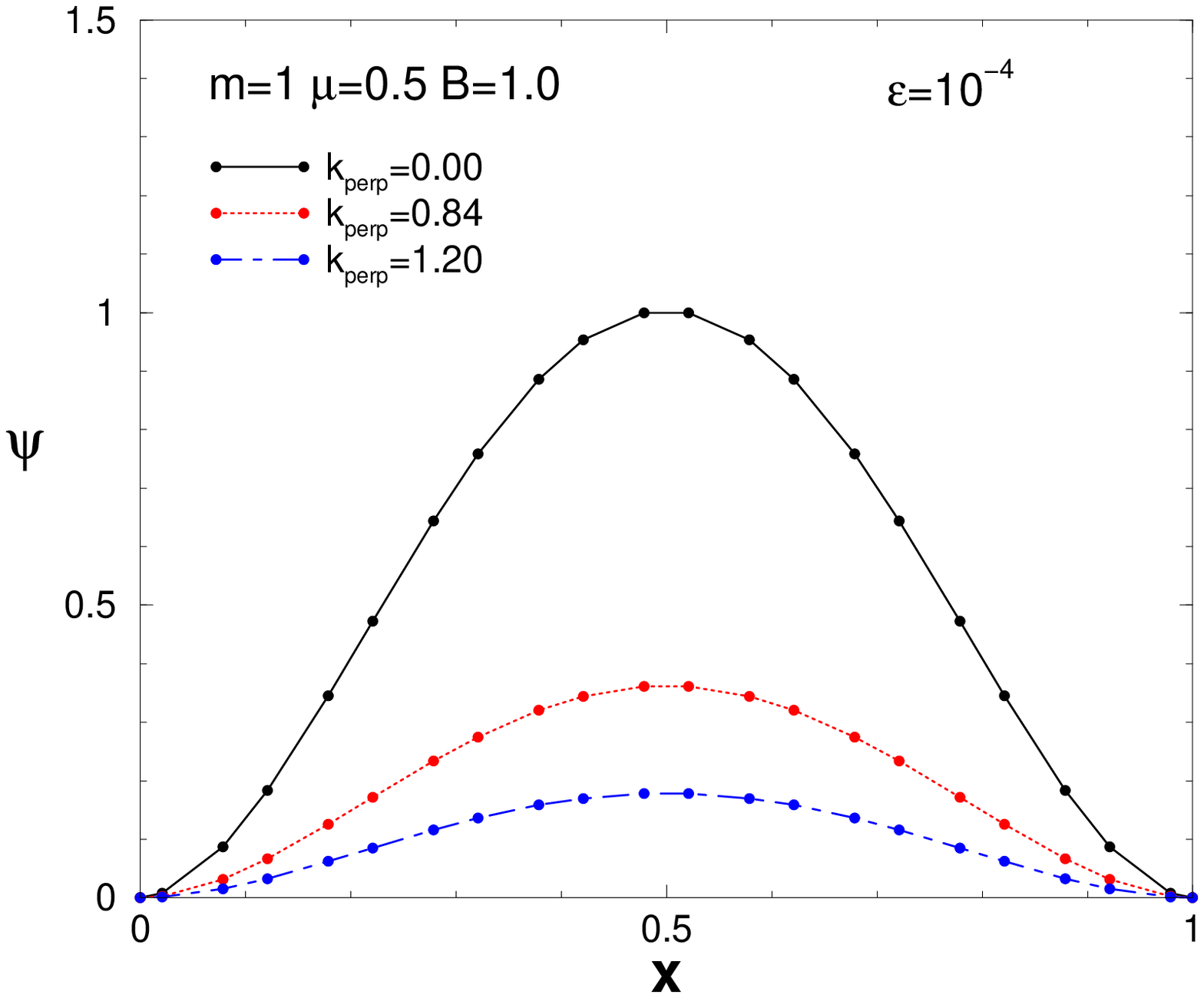}
 \vspace{-1cm}
\caption[*]{Wave function $\psi(k_{\perp},x)$ for ladder kernel
with $\mu=0.5$ and $B=1.0$. On left versus $k_{\perp}$ for fixed
values of $x$ and on right versus $x$ for a few fixed values of
$k_{\perp}$.} \label{psi_k}
\end{figure*}

The ladder BS amplitude at the rest frame $\vec{p}=0$ is shown in
Figure \ref{Phi_k}. On the left, we have plotted $\Phi(k_0,k)$ in
Minkowski space versus $k_0$. It exhibits a singular behaviour due
to the poles of the propagators in (\ref{bs}) at $k_0=\pm
\left(\varepsilon_k\pm \frac{M}{2}\right)$, {\it i.e.} poles
moving with $\vec{k}$ and $M$. On the right, we have plotted
$\Phi_E(k_4,k)=\Phi(k_0=ik_4,k)$ in Euclidean space versus $k_4$.
Both are calculated by eq. (\ref{bsint}) with the substitution
$k_0=ik_4$ for the Euclidean BS amplitude $\Phi_E(k_4,k)$. They
drastically differ from each other though they correspond to the
same binding energy. The Euclidean solution is smooth, in contrast
to the Minkowski one. The Euclidean amplitude found by direct
solution of the Wick-rotated equation (\ref{bs2}) is
indistinguishable from the one shown at r.h.-side of Figure
\ref{Phi_k}. The corresponding LF wave function
$\psi(k_{\perp},x)$ is shown in Figure \ref{psi_k}.

\begin{figure}[h!]
\begin{center}
\includegraphics[width=7.5cm]{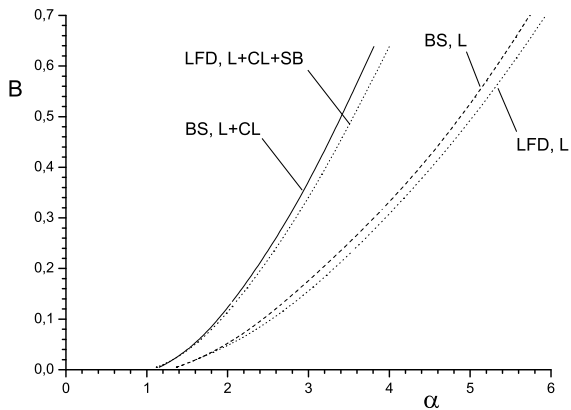}
\end{center}
 \vspace{-.7cm}
  \caption{Binding energy $B$ versus coupling constant
$\alpha$ for BS and LFD equations with ladder (L),
ladder+cross-ladder (L+CL), and ladder+cross-ladder+stretched box
(L+CL+SB) kernels. The exchange mass is $\mu=0.5$.}\label{figmu05}
 \vspace{-.7cm}
\end{figure}

The cross ladder effects are presented in Figure \ref{figmu05},
where the binding energy $B$ as a function of the coupling
constant $\alpha$ is shown for exchange masses $\mu=0.5$.
Corresponding numerical values --  with an accuracy of 1\% -- are
given in Table \ref{tab2}.
\begin{table*}[htbp]
\begin{center}
\caption{Coupling constant $\alpha$ for  given values of the
binding energy $B$ and exchanged mass $\mu=0.5$ calculated with BS
and LF equations for the ladder (L), ladder +cross-ladder (L+CL)
and (in LFD) for the ladder +cross-ladder +stretched-box (L+CL+SB)
kernels.}
\label{tab2}       
\begin{tabular}{cccccc}
\hline\noalign{\smallskip}
B & BS(L) & BS(L+CL)& LF(L) & LFD(L+CL)& LFD(L+CL+SB)\\
\noalign{\smallskip}\hline\noalign{\smallskip}
0.01 & 1.44 & 1.21 & 1.46 & 1.23 & 1.21 \\
0.05 & 2.01 & 1.62 & 2.06 & 1.65 & 1.62 \\
0.10 & 2.50 & 1.93 & 2.57 & 2.01 & 1.97 \\
0.20 & 3.25 & 2.42 & 3.37 & 2.53 & 2.47 \\
0.50 & 4.90 & 3.47 & 5.16 & 3.67 & 3.61 \\
1.00 & 6.71 & 4.56 & 7.17 & 4.97 & 4.91 \\
\noalign{\smallskip}\hline
\end{tabular}
\end{center}
\end{table*}
We see that for the same kernel -- ladder or ladder +
cross-ladder -- BS and LFD binding energies are very close to each other.
The BS equation is
slightly more attractive than LFD. At the same time, the results
for ladder and ladder+cross-ladder kernels considerably differ
from each other. The effect of the cross-ladder is strongly
attractive. Though the stretched box graphs are included, their
contribution to the binding energy is smaller than 2\% and
also attractive. This agrees with the direct calculation of the
stretched box contribution to the kernel performed in \cite{sbk}.
The solution of the Wick rotated equation (\ref{bs2}) for L+CL
kernel was found in \cite{mariane2}. Within a numerical accuracy
of 1\%, it coincides with the corresponding solution given by our
method, {\it i.e.}, with the column BS(L+CL) of the Table
\ref{tab2}. This coincidence proves the possibility to solve BS
equation with cross ladder kernel by using the Wick rotation. It
also confirms the validity of our method. We emphasize again the
fact that the Wick rotated equation allows to find the binding
energy, the same as the one determined by the Minkowski BS
equation, but not the Minkowski BS amplitude.

\section{CONCLUSION}\label{concl}

We have developed a new method for solving the Bethe-Salpeter
equation in the Minkowski space, {\it i.e.} without making use of
the Wick rotation. This method is based on an integral transform
of the original BS equation which removes its singularities.

Our equation is more easily formulated in terms of the weight
function of the Nakanishi integral representation \cite{nak63}.
This allows a straightforward reconstruction  of the original
Bethe-Salpeter amplitude -- both in Minkowski and Euclidean space
-- as well as the Light-Front wave function.

For zero-mass ladder exchange, we reproduce analytically  the
Wick-Cutkosky model. For massive ladder exchange, the binding
energies are found numerically and are in full agreement with
preceding results obtained in the Euclidean space.

For a given kernel, Bethe-Salpeter and Light-Front approaches give
very close results, though the first one is always slightly more
attractive. The cross-ladder contribution is strongly attractive
in both models. The  stretched-box contribution -- evaluated
separately only in the Light-Front framework -- is attractive but
small.

Binding energies obtained by solving Bethe-Salpeter
equation in  Minkowski space with ladder plus cross-ladder kernel
are also in full agreement with the ones obtained in Euclidean
space \cite{mariane2}.

We are aware about only one work \cite{ADT} where the separated
effect of the cross-ladder diagrams in the Bethe-Salpeter equation
has been estimated. The authors used an approximate dispersion
relation to obtain the corresponding kernel. Our results are found
to be three times smaller than those given in this reference.

Our method for solving the Bethe-Salpeter equation in the
Minkowski space can be generalized to non-zero angular momentum
and, presumably, to the fermion case.



\begin{thebibliography}{10}
\bibitem{SB_PR84_51}
E.E.~Salpeter, H.A.~Bethe, Phys. Rev. 84 (1951) 1232.

\bibitem{nakanishi} N.~Nakanishi, Prog. Theor. Phys. Suppl. 43 (1969) 1;
 95 (1988) 1.

\bibitem{WICK_54} G.C.~Wick,     Phys. Rev. 96 (1954) 1124.

\bibitem{nak63}
N.~Nakanishi, Phys. Rev.   130 (1963) 1230; {\it Graph Theory and
Feynman Integrals}, Gordon and Breach, New York, 1971.

\bibitem{KW} K.~Kusaka, A.G.~Williams, Phys. Rev.   D 51 (1995) 7026;
K.~Kusaka, K.~Simpson, A.G.~Williams, Phys. Rev.   D 56 (1997)
5071.

\bibitem{bbmst} S.G.~Bondarenko, V.V.~Burov, A.M.~Mo\-lo\-chkov,
G.I.~Smirnov and H.~Toki, Prog. in Part. and Nucl. Phys.,   48
(2002) 449.

\bibitem{sfcs}
J.H.O.~Sales, T.~Frederico, B.V.~Carlson and P.U.~Sauer,
Phys. Rev.   C 61 (2000) 044003;\\
T.~Frederico, J.H.O.~Sales, B.V.~Carlson and P.U.~Sauer, Nucl.
Phys.   A 737 (2004) 260.

\bibitem{bs1}
V.A. Karmanov and J. Carbonell, Eur. Phys. J. A, to be published;
arXiv:hep-th/0505261.

\bibitem{bs2}
J. Carbonell and V.A. Karmanov, Eur. Phys. J. A, to be published;
arXiv:hep-th/0505262.

\bibitem{cdkm}
J.~Carbonell, B.~Desplanques, V.A.~Karmanov and
\mbox{J.-F.}~Mathiot, Phys. Reports,   300 (1998) 215.

\bibitem{CUTKOSKY_PR96_54} R.E.~Cutkosky, Phys. Rev.   96 (1954) 1135.

\bibitem{mariane} M.~Mangin-Brinet and J.~Carbonell,
Phys. Lett.,   B 474 (2000) 237.

\bibitem{NT_FBS_96} T.~Nieuwenhuis and J.A.~Tjon, Few-Body Systems
21 (1996) 167.

\bibitem{sbk}
N.C.J.~Schoonderwoerd, B.L.G.~Bakker and V.A.~Karmanov, Phys. Rev.
{\bf C 58} (1998) 3093.

\bibitem{mariane2} M.~Mangin-Brinet and V.A.~Karmanov,
to be published.

\bibitem{ADT}
A.~Amghar, B.~Desplanques and L.~Theusl, Nucl. Phys.  A 694 (2001)
439.

\end{thebibliography}
\end{document}